\documentclass[aps,prc,twocolumn,superscriptaddress,showpacs]{revtex4}
\usepackage{epsfig}
\usepackage{dcolumn}
\begin{document}

\def\GSI{Gesellschaft f\"ur Schwerionenforschung mbH, D-64291 Darmstadt,
Germany}
\def\GANIL{GANIL, CEA et IN2P3-CNRS, F-14076 Caen, France}
\def\IPNO{Institut de Physique Nucl\'eaire, IN2P3-CNRS et Universit\'e, F-91406
Orsay, France}
\def\LPC{LPC, IN2P3-CNRS, ISMRA et Universit\'e, F-14050 Caen, France}
\def\SACLAY{DAPNIA/SPhN, CEA/Saclay, F-91191 Gif sur Yvette, France}
\def\LYON{Institut de Physique Nucl\'eaire, IN2P3-CNRS et Universit\'e, F-69622
Villeurbanne, France}
\def\NAPOLI{Dipartimento di Scienze Fisiche e Sezione INFN, Univ. Federico II,
I-80126 Napoli, Italy}
\def\CATANIA{Dipartimento di Fisica dell' Universit\`a and INFN, I-95129
Catania, Italy}
\def\WARSAW{A.~So\l{}tan Institute for Nuclear Studies, Pl-00681 Warsaw, Poland}
\def\MOSCOW{Institute for Nuclear Research, 117312 Moscow, Russia}
\def\IFJ{H. Niewodnicza\'nski Institute of Nuclear Physics, Pl-31342 Krak\'ow,
Poland}

\title{Transverse Velocity Scaling in $^{197}$Au + $^{197}$Au Fragmentation}

\affiliation{\GSI}
\affiliation{\GANIL}
\affiliation{\IPNO}
\affiliation{\LPC}
\affiliation{\SACLAY}
\affiliation{\LYON}
\affiliation{\NAPOLI}
\affiliation{\CATANIA}
\affiliation{\WARSAW}
\affiliation{\IFJ}
\affiliation{\MOSCOW}

\author{J.~{\L}ukasik}		\affiliation{\GSI}\affiliation{\IFJ}
\author{S.~Hudan}		\affiliation{\GANIL}
\author{F.~Lavaud}		\affiliation{\IPNO}
\author{K.~Turz\'o}		\affiliation{\GSI}
\author{G.~Auger}		\affiliation{\GANIL}
\author{Ch.O.~Bacri}		\affiliation{\IPNO}
\author{M.L.~Begemann-Blaich}	\affiliation{\GSI}
\author{N.~Bellaize}		\affiliation{\LPC}
\author{R.~Bittiger}		\affiliation{\GSI}
\author{F.~Bocage}		\affiliation{\LPC}
\author{B.~Borderie}		\affiliation{\IPNO}
\author{R.~Bougault}		\affiliation{\LPC}
\author{B.~Bouriquet}		\affiliation{\GANIL}
\author{Ph.~Buchet}		\affiliation{\SACLAY}
\author{J.L.~Charvet}		\affiliation{\SACLAY}
\author{A.~Chbihi}		\affiliation{\GANIL}
\author{R.~Dayras}		\affiliation{\SACLAY}
\author{D.~Dor\'e}		\affiliation{\SACLAY}
\author{D.~Durand}		\affiliation{\LPC}
\author{J.D.~Frankland}		\affiliation{\GANIL}
\author{E.~Galichet}		\affiliation{\LYON}
\author{D.~Gourio}		\affiliation{\GSI}
\author{D.~Guinet}		\affiliation{\LYON}
\author{B.~Hurst}		\affiliation{\LPC}
\author{P.~Lautesse}		\affiliation{\LYON}
\author{J.L.~Laville}		\affiliation{\GANIL}
\author{C.~Leduc}		\affiliation{\LYON}
\author{A.~Le~F\`evre}		\affiliation{\GSI}
\author{R.~Legrain}		\affiliation{\SACLAY}
\author{O.~Lopez}		\affiliation{\LPC}
\author{U.~Lynen}		\affiliation{\GSI}
\author{W.F.J.~M\"uller}	\affiliation{\GSI}
\author{L.~Nalpas}		\affiliation{\SACLAY}
\author{H.~Orth}		\affiliation{\GSI}
\author{E.~Plagnol}		\affiliation{\IPNO}
\author{E.~Rosato}		\affiliation{\NAPOLI}
\author{A.~Saija}		\affiliation{\CATANIA}
\author{C.~Sfienti}		\affiliation{\GSI}
\author{C.~Schwarz}		\affiliation{\GSI}
\author{J.C.~Steckmeyer}	\affiliation{\LPC}
\author{G.~T\v{a}b\v{a}caru}	\affiliation{\GANIL} 
\author{B.~Tamain}		\affiliation{\LPC}
\author{W.~Trautmann}		\affiliation{\GSI}
\author{A.~Trzci\'{n}ski}	\affiliation{\WARSAW}
\author{E.~Vient}		\affiliation{\LPC}
\author{M.~Vigilante}		\affiliation{\NAPOLI}
\author{C.~Volant}		\affiliation{\SACLAY}
\author{B.~Zwiegli\'{n}ski}	\affiliation{\WARSAW}
\author{A.S.~Botvina}		\affiliation{\GSI}\affiliation{\MOSCOW}
\collaboration{The INDRA and ALADIN Collaborations}
\noaffiliation

\date{\today}

\begin{abstract}

Invariant transverse-velocity spectra of intermediate-mass fragments
were measured with the 4$\pi$ multi-detector system INDRA for
collisions of $^{197}$Au on $^{197}$Au at incident energies between
40 and 150 MeV per nucleon.                                             
Their scaling properties as a function of incident energy               
and atomic number $Z$ are used to distinguish and characterize
the emissions in (i) peripheral collisions at the projectile and
target rapidities, and in (ii) central and (iii) peripheral
collisions near mid-rapidity.
The importance of dynamical effects is evident in all three cases
and their origin is discussed.

\end{abstract}


\maketitle

Heavy-ion collisions in the Fermi-energy domain are rich and complex
sources of fragment emissions \cite{hirsch99}.
With heavy nuclei, rather large composite systems may be formed
\cite{dago96,marie97,frank01} that represent the limit in charge and mass
over which the bulk properties of excited nuclear matter may be
experimentally explored with present-day means.
More peripheral collisions may give access to systems consisting
of low-density matter interacting with two residual nuclei. 
Such configurations are of interest for the study of
isotopic effects in the phase behavior of the two-component nuclear fluid
\cite{muell95,shi00,poggi01,baran02}.

The fragment channels of both central and peripheral collisions 
exhibit characteristic dynamical phenomena. 
A collective flow, predominantly at small impact
parameters, is superimposed on the thermal motion of the emitted particles
and fragments, increasing in strength with the bombarding energy
\cite{reis97}. Peripheral collisions appear binary in the light-particle
channels dominated by evaporation from the projectile and target remnants
but are characterized by a strong component of fragment emission at
rapidities intermediate between those of the projectile
and of the target 
\cite{mont94,toke95,demp96,luka97,plagnol00,dore01,piant02}.
A possibly dynamical origin of this component, frequently termed neck
emission, has been suggested but its 
nature has not yet been satisfactorily clarified.

A well founded understanding of the mechanisms of fragmentation processes
is indispensable if their potential for investigating the thermodynamic
behavior of nuclear matter is to be exploited.  Considerable progress in
this direction has been made possible by the advent of 4$\pi$-type
detection systems that permit the construction of complete
Galilei or Lorentz invariant cross-section distributions.
It will be shown, in this Letter, that transverse-velocity spectra
obtained from such distributions are particularly useful for identifying
the qualitatively different types of emissions on the basis of
their characteristic dependences on impact parameter and incident
energy. A striking invariance with the incident energy is observed
for mid-rapidity fragments from peripheral collisions, the fragment 
component whose origin is probably the least-well understood.

The data were obtained with the INDRA multidetector
\cite{pouthas} in experiments performed at the GSI
for the reaction $^{197}$Au + $^{197}$Au
at incident energies from 40 to 150 MeV per nucleon, i.e. over a range of
relative velocities from once to twice the Fermi value.
Beams of $^{197}$Au delivered by the heavy-ion synchrotron SIS
were directed onto $^{197}$Au targets of 2-mg/cm$^2$ areal thickness.
Annular veto detectors directly upstream
of the INDRA detection system and measurements with empty target frames
have been
employed in order to verify that the synchrotron beams were properly
focussed into the 12-mm-diameter entrance hole of the 4$\pi$
detection device.
 
The energy calibration of the 336 CsI(Tl) detectors has primarily been 
derived from a detector-by-detector comparison of spectra remeasured for 
the reaction $^{129}$Xe + $^{nat}$Sn at 50 MeV per nucleon with spectra
measured previously at GANIL \cite{marie97,luka97,plagnol00}. There,
calibration data had been obtained by scattering a variety of primary and 
secondary beams from thin targets. The light emission from the CsI(Tl)
scintillators 
was parameterized according to Ref. \cite{parlog02}, and the $Z$ dependence
of the parameters was obtained from comparing the generated 
$\Delta E - E$ maps to predictions of energy-loss and range 
tables \cite{hubert}. 

As in previous investigations of symmetric heavy ion
collisions \cite{luka97,plagnol00}, the total 
transverse energy $E^{12}_{\perp}$ of light charged particles ($Z \le$ 2)
has been used as an impact parameter selector.
The $E^{12}_{\perp}$ spectra are found to scale linearly with the
incident energy,
and the scaled spectra coincide. The relation 
between $E^{12}_{\perp}$ and the reduced impact parameter $b/b_{\rm max}$,
obtained with the use of the geometrical prescription~\cite{cavata90},
is linear in very good approximation, where $b/b_{\rm max}$ decreases with
increasing $E^{12}_{\perp}$.
This behavior is very similar to that observed for the $^{129}$Xe + 
$^{nat}$Sn reaction at the lower energies 25 to 50 MeV per nucleon
\cite{luka97,plagnol00}. With the same prescription as used there,
impact parameter bins were generated. 
The most central bin, labelled bin 8,
covers impact parameters up to 5\% of $b_{\rm max}$. 
The remaining part of the $E^{12}_{\perp}$ spectrum is 
divided into 7 bins of equal width, corresponding to 7 bins of 
approximately equal width in $b$. Bin 1 contains the most 
peripheral collisions that were registered at the trigger condition 
of at least three detected particles.

\begin{figure}
     \epsfxsize=7.0cm
     \centerline{\epsffile{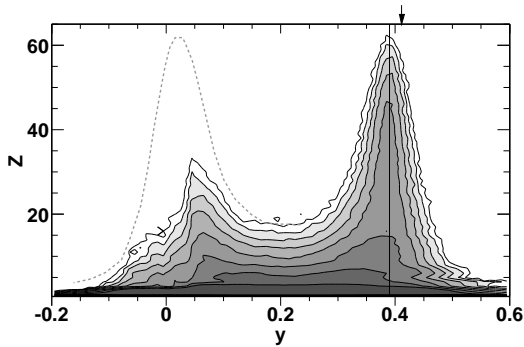}}
\caption{\label{fig:rap}
Contour plot of the measured production cross sections as a function
of the fragment $Z$ and the rapidity $y$ for the reaction 
$^{197}$Au + $^{197}$Au at 80 MeV per nucleon for semi-peripheral 
collisions (bin 3). 
The vertical line indicates the mean rapidity of heavy 
projectile-like 
fragments which is slightly below the projectile rapidity 
$y$ = 0.41 (arrow).
The dashed line is meant to visually restore the symmetry of the reaction 
which is disturbed by the reduced acceptance for low-energy fragments.
}
\end{figure}

\begin{figure}
     \epsfxsize=8.6cm
     \centerline{\epsffile{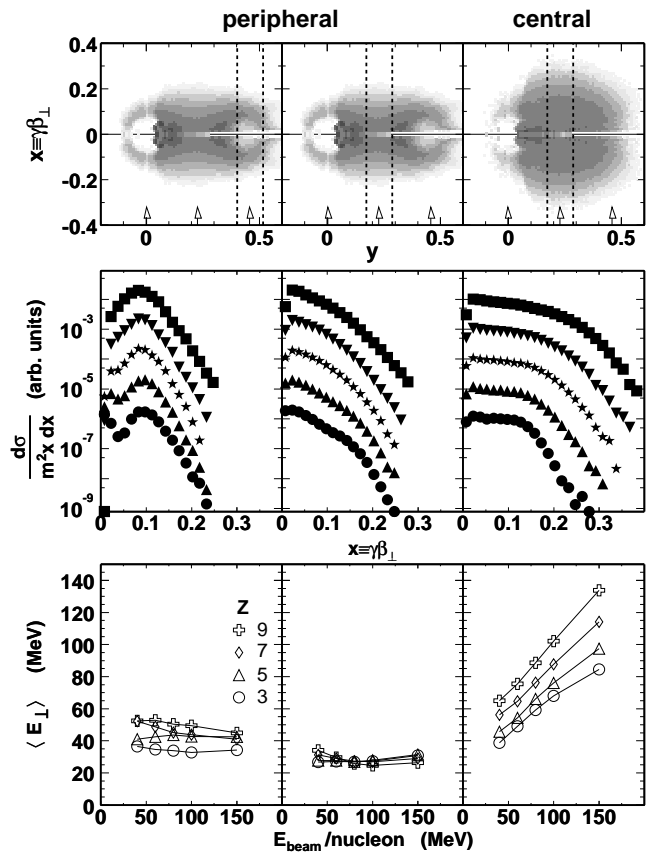}}
\caption{\label{fig:main}
Top row: invariant cross section distributions for $Z$ = 3 fragments 
as a function of transverse velocity ($x$) and rapidity ($y$) for 
peripheral (bin 1, left and center) and central 
(bins 7 and 8, right)
collisions of $^{197}$Au + $^{197}$Au at $E/A$ = 100 MeV. 
The dashed lines indicate the windows in relative rapidity chosen 
for the data shown below; the arrows denote the rapidities of
the target, the center-of-mass, and the incident projectile 
(from left to right). \\
Middle row: 
invariant transverse velocity spectra
for $Z$ = 3 at bombarding energies $E/A$ = 40, 60, 80, 100 and 150 MeV
(from bottom to top), vertically displaced for clarity and each 
plotted over three decades.\\
Bottom row: mean transverse energies $\langle E_{\perp} \rangle$
as a function of the incident energy for fragments with odd
$Z \le$ 9 as indicated.
}
\end{figure}

Peripheral collisions display a binary character in that 
heavy residues of the projectile and target are observed, 
as far as they are within the INDRA acceptance (Fig. \ref{fig:rap}).
The emission of intermediate-mass fragments, however, is neither binary
nor isotropic with respect to these residues 
(Figs. \ref{fig:rap},\ref{fig:main}).
Their intensity is heavily weighted towards mid-rapidity  
where, at the lower incident energies, it displays a pronounced maximum, 
similar as observed for $^{129}$Xe + $^{nat}$Sn \cite{luka97,plagnol00}. 
At higher incident energies, 100 MeV per nucleon and above,
the fragment distributions start to separate at mid-rapidity, and a 
cross-section minimum develops (Fig. \ref{fig:main}, top).
This may be seen as the beginning of a smooth evolution into the 
relativistic regime where the fragment emission is concentrated
at the projectile and target rapidities \cite{schuett96}.

As a potentially useful observable for identifying the dominant mechanisms
of fragment production, invariant transverse-velocity spectra 
for selected centrality and rapidity bins
are presented in Fig. \ref{fig:main}. 
These spectra are expected to be Gaussian for a thermally emitting source,
with a width $\sigma^2 = T/m$ where $T$ and $m$ are the temperature of the 
source and the mass of the emitted particle, respectively. If
Coulomb forces act in addition, a peak will appear near the velocity
corresponding to the Coulomb energy.

The cuts in rapidity $y$ that were used to generate the spectra are 
indicated in the invariant cross section plots, shown for $Z$ = 3
and 100 MeV per nucleon
in the upper row of Fig.~\ref{fig:main}. Identical cuts in the
scaled rapidity $y/y_{p}$, with a width of 25\% of $y_{p}$, were used for 
the other bombarding energies ($y_{p}$ is the projectile rapidity).
Impact-parameter bin 1 was chosen 
to represent peripheral collisions, and bins 7 and 8 were combined to 
obtain adequate statistics for central collisions. The transverse-velocity 
spectra of Li nuclei for the five bombarding energies are shown in the middle 
panels of the figure. The mean transverse energies for fragments with
atomic number $Z$ = 3, 5, 7, 9 as a 
function of the incident energy are given below.

The transverse velocities near the projectile rapidity in peripheral 
collisions (left column) are dominated by a prominent 
Coulomb peak that indicates repulsion from
the surface of a heavy primary fragment. 
The peak velocity $\beta \approx$ 0.09, equivalent to 2.7 cm/ns
and typical for fragment emission from a gold-like residue \cite{trock87},
is rather stable and drops by only
$\Delta\beta \approx$ 0.01 as the energy is raised from 40
to 150 MeV per nucleon.

Coulomb peaks are absent in the emissions at midrapidity which exhibit two
different scaling behaviors for the central and peripheral impact 
parameters. In the central case (Fig. \ref{fig:main}, right panel), 
the shapes are approximately 
Gaussian, with an extra shoulder superimposed at the lower incident 
energies, most likely due to Coulomb repulsion. 
Both, the mean velocity and the 
width increase considerably with increasing bombarding energy and with 
the fragment mass. The mean transverse energies, correspondingly,
grow approximately linearly with the incident energy but slightly slower 
than in proportion to the fragment mass (cf. Ref. \cite{hsi94}). 
These observations reflect the increasing collectivity
of the fragment motion as the incident energy rises, a result of higher 
compression, a resulting stronger Coulomb acceleration, 
and higher temperatures 
of the composite sources that are initially formed 
in central collisions \cite{nebau99,inpc}.

The most striking behavior is observed for the mid-rapidity fragments 
from peripheral reactions (middle column of Fig. \ref{fig:main}).
The shapes of the transverse-velocity spectra, somewhat between Gaussian and
exponential, are virtually the same at all incident energies,
except perhaps for a weak shoulder that develops at the lowest energies 
(middle panel). The mean transverse energies are invariant with respect to 
the incident energy and to the fragment $Z$ (bottom panel). 
The existence of a common energy scale 
for intermediate-mass fragments over the investigated wide range 
of bombarding energies seems, at first sight,
unexpected for a dynamical mechanism and raises the 
question of where to search for its origin.

\begin{figure}
     \epsfxsize=7.0cm
     \centerline{\epsffile{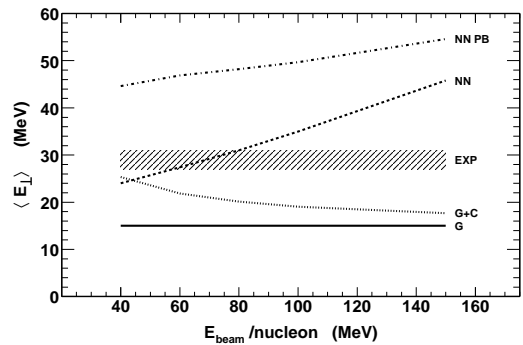}}
\caption{\label{fig:last}
Mean transverse energies $\langle E_{\perp} \rangle$ as obtained from the
Goldhaber model for $p_{\rm F}$ = 265 MeV/c (full line labelled G), 
after adding the Coulomb energy (dotted line, G+C), and 
for nucleons from primary nucleon-nucleon collisions with (dashed-dotted 
line, NN PB) and without (dashed line, NN) considering Pauli blocking.
The experimental result for midrapidity fragments from peripheral collisions
is represented by the hatched area.
}
\end{figure}

Kinetic energies that are independent of the particle species are expected 
within thermal models. In the present case, however, a mean transverse 
energy $\langle E_{\perp} \rangle \approx$ 30 MeV, 
corresponding to a temperature $T$ of the same 
magnitude, seems rather large and clearly exceeds the temperature range at 
which fragments can be expected to survive. Kinetic energies 
that appear thermal and correspond to high temperatures are obtained from
the Goldhaber model in which 
fragment momenta are assumed to result from the nucleonic Fermi 
motion \cite{gold74}. This approach has proven useful for the 
interpretation of kinetic energies of intermediate-mass fragments from
spectator decays at 
relativistic bombarding energies \cite{odeh00}. In the present case,
however, the value 
$\langle E_{\perp} \rangle$ = $T$ = 15 MeV, derived for the Fermi 
momentum $p_{\rm F}$ = 265 MeV/c of heavy nuclei, amounts to only about 
half of the equivalent temperature that is required (Fig. \ref{fig:last}).
Since it is 
unlikely that the density or the temperature of the neck region attain the
extreme values that would be needed to bring this value up and close to 
observation \cite{bauer95}, at most part of the observed energies can be 
ascribed to the intrinsic nucleonic motion. 

The intermediate bombarding energies are characterized by the interplay of
mean-field and nucleon-nucleon collision dynamics. The importance 
of the latter increases with energy, due to the reduced role of Pauli 
blocking, as apparent in the prompt emissions 
of light particles \cite{lemmon99,dore01}. 
The transverse energies generated in primary nucleon-nucleon collisions
should reflect the incident energy, in addition to the Fermi motion. 
The rise with energy, however, will be partly compensated by the reduced
Pauli blocking of final states with low transverse momentum at higher
incident energies.
Simulations confirm that essentially only the widths of the distributions 
increase while the mean transverse 
energies rise rather slowly if the blocking effect is taken into account
(Fig. \ref{fig:last}). This property of the mean transverse energies 
is specific for the present range of energies at and above the 
Fermi value. Fragments formed by 
a coalescence mechanism from only scattered nucleons will have transverse 
energies that are clearly too high. It will be sufficient if a few of 
them are built into the nascent fragments which may also trigger
their separation from the bulk.

The stronger Pauli blocking at the lower incident energies thus 
represents a compensating mechanism that contributes to the observed 
invariance of $\langle E_{\perp} \rangle$.
This is also the case for the Coulomb potential 
generated by the two residues in the neck region.
Essential for reactions at lower energies and in fission 
\cite{poggi01,piant02,field92}, its role is expected to decrease at 
the higher bombarding energies at which the residue velocities are
comparable to or larger than those of the mid-rapidity fragments. 
Simulations with Coulomb trajectory calculations
confirm this effect (Fig. \ref{fig:last}). They also indicate 
that the Coulomb contribution to the transverse energy may depend 
very little on the fragment $Z$. In the simulations, a thermal 
distribution was assumed for the initial fragment motion, in accordance 
with the Goldhaber model. It has the consequence 
that heavier fragments move away more slowly from the field free zone in 
between the receding residues. This may explain why the
Coulomb forces do not disturb the $Z$ invariance of the transverse motion
(Fig. \ref{fig:main} bottom, middle panel).

All these mechanisms, whose effects are schematically illustrated in
Fig. \ref{fig:last}, are contained in realistic microscopic transport
theories which thus seem suited to test the suggested reason for
the observed invariances \cite{wada00}.
Complications may arise from the fact that surface effects and finer details 
in the modelling are important 
in peripheral reactions and will require a rather careful treatment
\cite{plagnol00,aichelin}. 
Additional insight, in particular with regard to the role of the 
Coulomb forces, may also be gained from an experimental study of the 
dependence of the fragment transverse motion on the mass of the 
collision system.

To summarize, three different modes of fragment emission have been 
characterized on the basis of the transverse-velocity spectra. All 
of them are strongly influenced by dynamical effects. Peripheral collisions
appear binary only as long as the heavy residues and evaporated 
light particles are considered. The emission of intermediate-mass 
fragments is not symmetric with respect to the residue rapidities. The
mid-rapidity fragments from these collisions exhibit large transverse 
velocities on a scale corresponding to an effective temperature of 
nearly 30 MeV. The invariance of their transverse motion with incident 
energy and fragment $Z$ is most likely the result of a compensation of 
several dynamical effects, the initial Fermi motion in the colliding 
nuclei, the generation of transverse momenta in nucleon-nucleon 
collisions and the Coulomb interaction between the fragments and the 
separating residues. 
The important role of collective motion, linearly increasing with the 
incident energy over the covered range, is evident for central collisions.

The authors would like to thank the staff of the GSI for 
providing heavy ion beams of the highest quality and for technical support.
M.B. and C.Sc. acknowledge the financial support
of the Deutsche Forschungsgemeinschaft under the Contract No. Be1634/1-
and Schw510/2-1, respectively; D.Go. and C.Sf. acknowledge the receipt of 
Alexander-von-Humboldt fellowships.
This work was supported by the European Community under
contract ERBFMGECT950083.

\end{document}